\RequirePackage{fixltx2e}
\documentclass[notitlepage,twocolumn,prd,floatfix,showpacs,aps]{revtex4-1}
\usepackage[nodayofweek]{datetime}
\usepackage{paralist}
\usepackage{amsmath}
\usepackage{mathtools}
\usepackage{natbib}
\usepackage{color}
\usepackage{nicefrac}
\usepackage{hyperref}
\usepackage{graphicx}
\usepackage[caption=false]{subfig}
\newcommand{\kms}{\,\textrm{km s}^{-1}}

\usepackage{multirow}

\begin{document}

\title{Improved determination of the WIMP mass from direct detection data}

\author{Bradley J. Kavanagh}
\email{Electronic address: ppxbk2@nottingham.ac.uk}
\author{Anne M. Green}
\email{Electronic address: anne.green@nottingham.ac.uk}
\affiliation{School of Physics \& Astronomy,University of Nottingham, University Park, Nottingham, NG7 2RD, UK}

\date{\today}

\begin{abstract}
Direct detection experiments searching for weakly interacting massive particle (WIMP) dark matter typically use a simplified model of the Galactic halo to derive parameter constraints. However, there is strong evidence that this Standard Halo Model is not a good approximation to our Galaxy. We discuss previous attempts to extract the WIMP mass, cross-section and speed distribution from direct detection data and show that these lead to significant biases in the reconstructed parameter values. We develop and test an alternative model-independent method based on parametrising the \textit{momentum} distribution of the WIMPs. This allows us to limit the analysis only to those regions of momentum space to which the experiments are sensitive. The method can be applied to a single experiment to extract information about the distribution function, as well as information on the degenerate WIMP mass and interaction cross-section combined in a single parameter. This degeneracy can be broken by including data from additional experiments, meaning that the WIMP mass and speed distribution can be recovered. We test the momentum parametrisation method using mock datasets from proposed ton-scale direct detection experiments, showing that it exhibits improved coverage properties over previous methods, as well as significantly reduced bias. We are also able to accurately reconstruct the shape of the WIMP speed distribution but distinguishing between different underlying distributions remains difficult.
\end{abstract}

\pacs{07.05.Kf,14.80.-j,95.35.+d,98.62.Gq}

\maketitle

\section{Introduction}

A large abundance of particle dark matter (DM) is required in our Universe to explain many observations on scales from the galactic to the cosmological (for a review, see e.g.\ Ref.\ \cite{DMO8}). Thus far, these particles have only been observed via their gravitational interactions and there are many experiments - completed, in progress and in development - which aim to directly detect Galactic dark matter through its non-gravitational interactions in the laboratory. While there is no lack of well-motivated DM candidates (e.g.\ Ref.\ \cite{DMAX8,DMN1,DMO12,ADM1}), the nature of the dark matter particles remains an open question. Different classes of experiment are sensitive to different candidates and it is necessary to rely on a wide range of experiments to constrain the exact properties of the dark matter.

We focus here on the search for weakly interacting massive particle (WIMP) \cite{DMO13} dark matter. WIMP direct detection experiments aim to identify and measure the energies of nuclear recoils induced by WIMP-nucleon collisions \cite{DMO9,DMO11}. Excesses above the expected backgrounds have been observed in the CoGeNT \cite{DMDD5, DMDD12} and CRESST-II \cite{DMDD59} experiments. In addition, an annual modulation in the event rate has been observed at \(8.9\sigma\) significance by the DAMA/LIBRA collaboration \cite{DMDD4}. While these tentative signals have been cited as evidence for a light WIMP of mass \(\sim 10 \textrm{ GeV}\), the various results are difficult to reconcile in simple models of dark matter \cite{DMDD17,DMDD92,DMDD31} and are in significant tension with the null results presented by CDMS \cite{DMDD63} and XENON100 \cite{DMDD6}. It is important to accurately extract the properties of the WIMP from such experiments in order to check consistency with other detection channels (such as indirect detection \cite{DMID21} and constraints from colliders \cite{DMCol3}), as well as to provide constraints on the underlying models of supersymmetry \cite{DMO4}. 

WIMP direct detection experiments are typically analysed within the framework of the Standard Halo Model (SHM) in which the velocity of galactic dark matter particles is assumed to follow a Maxwell-Boltzmann distribution. This is a highly simplified assumption and can lead to a bias in the values of the WIMP mass and cross-section measured in direct detection experiments (see Ref.\ \cite{DMDD49,DMDD19} and references therein). 

There have been several attempts to develop model independent analysis schemes in order to avoid this bias as well as directly measuring the WIMP velocity distribution. Frandsen et al.\ \cite{DMDD19,*DMDD98} attempt to factorise out the astrophysical uncertainties, allowing a upper bound to be placed on the WIMP cross-section, while Alves et al.\ \cite{DMDD85} parametrise the velocity distribution in terms of Legendre polynomials and attempt to reconstruct the polynomial coefficients from directional experiments. However, these techniques must be applied at a given WIMP mass, which would need to be measured independently (for example, in a collider experiment). Drees \& Shan \cite{DMDD20,DMDD97} have proposed calculating moments of the velocity distribution function directly from data and using this to obtain the WIMP mass. However, the inclusion of finite energy windows in the mock experiments leads to significant bias and unreliable error estimates. Strigari \& Trotta \cite{DMDD39} suggest using a simplified model of the Milky Way halo and simultaneously fitting the WIMP and halo parameters. However, the effectiveness of this method depends on how closely the actual Galactic halo can be described by such a model. 

Here, we develop further the methods of Peter \cite{DMDD2,DMDD3} (see also Ref.\ \cite{DMDD48}), who has attempted to reconstruct the WIMP mass, cross-section and distribution function simultaneously by using a simple empirical parametrisation of the velocity distribution. The analysis of mock datasets using this method indicates a significant bias in the reconstructed parameters even for simple velocity distributions. We present an alternative method based on parametrising the \textit{momentum} distribution of the WIMPs, which allows us to probe only that region of parameter space to which the experiment is sensitive. We test the method using mock datasets from proposed ton-scale direct detection experiments to assess its effectiveness.

Section \ref{sec:Framework} of this paper presents the current framework under which direct detection data is analysed, highlighting the need for alternative methods. Sec.\ \ref{sec:Parameters} details the Markov Chain Monte Carlo techniques which will be used to assess the proposed model-independent methods, as well as the benchmark parameters used to generate mock datasets.  The statistical properties of previous speed parametrisation methods will be explored in Sec.\ \ref{sec:SpeedMethod}. We then present the new proposed momentum parametrisation method and how it can be applied to a single experiment (Sec.\ \ref{sec:MomentumMethod1}) and to combined datasets from several experiments (Sec.\ \ref{sec:MomentumMethod2}). The results of large numbers of parameter reconstructions are also presented in these Sections, before moving on to summarise the significance of this study in Sec.\ \ref{sec:Conclusion}.

\section{Direct Detection Analysis}
\label{sec:Framework}
The differential event rate per unit detector mass for nuclear recoils of energy \(E_R\) due to WIMP elastic scattering is given by \cite{DMO4}:

\begin{equation}
\frac{\textrm{d}R}{\textrm{d}E_R} = \frac{\rho_0 \sigma_p}{2 \mu_{\chi p}^2 m_\chi} A^2 F^2(E_R) \eta(v_{\textrm{min}})\,,
\end{equation}
where \(\rho_0\) is the local DM mass density, \(m_\chi\) is the WIMP mass, \(\sigma_p\) is the DM-proton cross-section (at zero-momentum transfer) and \(\mu_{\chi p}\) is the WIMP-proton reduced mass, \(\mu_{\chi p} = (m_\chi m_p)/(m_\chi + m_p)\). Here, we have considered only the spin-independent contribution to the rate, which dominates for targets with large atomic mass number, \(A\). Assuming that the coupling to protons and neutrons is identical, this leads to an \(A^2\) enhancement in the scattering rate \cite{DMO4}. Finally, the loss of coherence due to the finite size of the nucleus is accounted for by the form factor, \(F^2(E_R)\), which we take to have the Helm form \cite{Helm}.

The local velocity distribution of galactic WIMPs enters into the event rate through \(\eta(v_{\textrm{min}})\), the mean inverse speed:

\begin{equation}
\eta(v_{\textrm{min}}) = \int_{v_{\textrm{min}}}^\infty \frac{f(\textbf{v})}{v}\, \textrm{d}^3\textbf{v} \,,
\end{equation}
where \(f(\textbf{v})\) is the normalised velocity distribution in the Earth frame and \(v_{\textrm{min}}\) is the minimum WIMP speed required to produce a recoil of energy \(E_R\). For a detector of nuclear mass, \(m_N\), and reduced WIMP-nucleon mass, \(\mu_{\chi N} = (m_\chi m_N)/(m_\chi + m_N)\), this minimum required speed is

\begin{equation}
v_{\textrm{min}}(E_R, m_\chi, m_N) = \sqrt{\frac{m_N E_R}{2\mu_{\chi N}^2}} \,.
\end{equation}

Each experiment is then sensitive to a particular range of WIMP speeds, \(v \in \left[ v_{\textrm{min}}(E_{\textrm{min}}), v_{\textrm{min}}(E_{\textrm{max}})\right]\), where \(E_\textrm{min}\) and \(E_\textrm{max}\) are the lower and upper limits of the signal window respectively. WIMPs with speeds below \(v_{\textrm{min}}(E_{\textrm{min}})\) do not have sufficient kinetic energy to induce nuclear recoils above the threshold energy of the detector. WIMPs with speeds above \(v_{\textrm{min}}(E_{\textrm{max}})\) will contribute to recoils in the detector, but this contribution will be a constant offset to the rate and as such the experiment will not be sensitive to the exact form of the speed distribution above this speed.

Data from direct detection experiments are typically analysed using the Standard Halo Model (SHM), in which the galactic halo is modelled as an isothermal sphere with Maxwellian velocity distribution:

\begin{equation}
f(\textbf{v}) = \frac{1}{(2\pi)^{\nicefrac{3}{2}} \sigma^3} \, \exp \left(-(\textbf{v} - \textbf{v}_\textrm{lag}) ^2/2\sigma^2\right).
\end{equation}
A typical value for the velocity dispersion is \(\sigma \approx 156 \kms\), in order to match the peak speed to the local circular speed: \(v_0 = \sqrt{2}\sigma \approx 220 \kms\).  The motion of the Earth with respect to the Galactic halo is encoded in \(\textbf{v}_\textrm{lag}\), which incorporates the Sun's motion within the Galaxy and the Earth's orbital motion about the Sun. Here, we neglect the time variation of \(\textbf{v}_\textrm{lag}\) and use an average value of \(|\textbf{v}_\textrm{lag}| \approx 230 \kms\) \cite{DMO4}.

We consider only non-directional detectors, for which it is useful to define the directionally averaged local \textit{speed} distribution, \(f(v)\):
\begin{equation}
f(v) = \oint f(\textbf{v}) \, \textrm{d}\Omega_\textbf{v}.
\end{equation}
We also define the `3-dimensional' (3-D) speed distribution, \(f_3(v) = v^2 f(v)\), such that \(f_3(v) \,\textrm{d}v\) is the fraction of WIMPs with speeds between \(v\) and \(v + \textrm{d}v\).

Despite the Standard Halo Model's frequent use, there is strong evidence that the true distribution function of galactic WIMPs is non-Maxwellian. Even assuming the Standard Halo Model, there remains a great deal of uncertainty in the correct parameter values which should be used (see e.g.\ \cite{DMA49,DMA50}). In addition, high resolution N-body simulations suggest that there can be significant deviations from the maxwellian velocity distribution \cite{DMA43, DMA52,DMA53}, including features in \(f(\textbf{v})\) especially at high velocity. Some simulations also show evidence for a dark disk, corotating with the baryonic disk of the Milky Way, caused by the tidal distruption of satellites which are preferentially dragged into the disk plane \cite{DMA54,DMA55}. It is therefore necessary to consider the presence of additional velocity structure beyond the SHM and the impact this may have on parameter reconstruction.

\section{Parameter Estimation Method}
\label{sec:Parameters}
We now present the benchmark experimental and theoretical parameters which will be used to test various reconstruction methods as alternatives to the simple SHM assumption. The Markov Chain Monte Carlo methods used to explore the posterior distribution of the parameters are also introduced. 

\subsection{Benchmark Parameters and Experiments}
\label{sec:ParamBenchmarks}
As noted in Ref.\ \cite{DMDD97}, it is impossible to estimate the WIMP mass from a single experiment if no assumptions are made about \(f(v)\), so we consider three next-generation detectors, modelled on experiments which are currently in development. Each experiment is characterised by a single (suitably averaged) target nucleus mass, \(m_N\), a total detector mass, \(m_\textrm{det}\), an effective exposure time, \(t_\textrm{exp}\), and a pair of energies, \(E_\textrm{min}\) and \(E_\textrm{max}\), which mark the extent of the signal region. We focus on a particular set of experimental parameters in order to provide a concrete example of how the WIMP parameters can be estimated accurately. Table \ref{tab:Expts} shows the experimental parameters used in this work, which are chosen to approximately match those used by Peter \cite{DMDD3}.

\begin{table*}[t]
  \begin{center}
\begin{ruledtabular}
    \begin{tabular}{llll}

    & XENON1T \cite{DMDD95} & SuperCDMS \cite{DMDD74} & WArP \cite{DMDD94} \\
\hline
Detector Target & Xe & Ge & Ar \\
Nuclear Mass, \(m_\textrm{N} / \textrm{amu}\) & 131 & 73 & 40 \\
Detector Mass, \(m_\textrm{det} / \textrm{kg} \) & 1000  & 100 & 1000 \\
Exposure Time, \(t_\textrm{exp} / \textrm{days} \) & 60.8  & 109.5 & 365\\
Energy Range, \(\left[E_\textrm{min},E_\textrm{max} \right] / \textrm{keV}\) & [2,30] & [10,100] & [30,130] \\
    \end{tabular}
\end{ruledtabular}
  \end{center}
  \caption{Parameter values for the three mock experiments used in this work, chosen to closely match those used in Ref.\ \cite{DMDD3}. The meanings of the experimental parameters are described in Sec.\ \ref{sec:ParamBenchmarks}.}
\label{tab:Expts}
\end{table*}

We assume perfect and uniform detection efficiency - that is, all signal events and no background events survive analysis cuts. We also assume perfect energy resolution. For a real experiment, these assumptions will almost certainly not hold, for example due to variations in the relative scintillation efficiency of Xenon \cite{DMDD89}, but the results presented here should be viewed as a proof of principle in the ideal case.

\begin{figure}[t]
  \centering
  \includegraphics[width=0.5\textwidth]{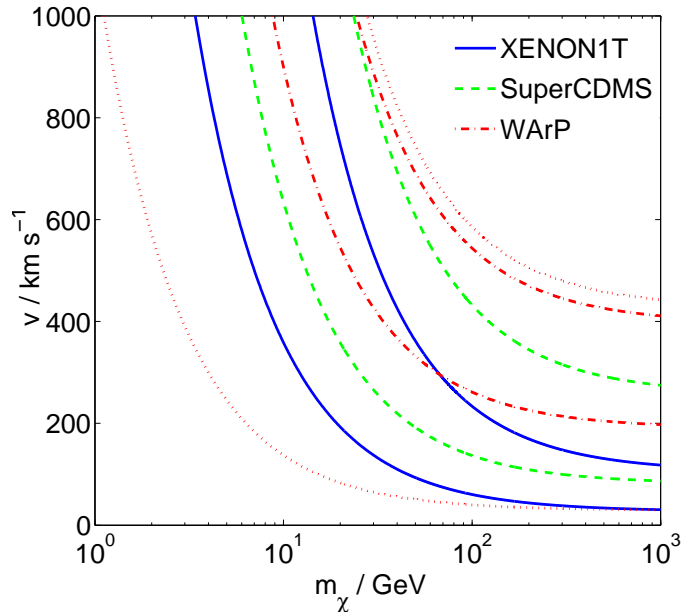}
\caption{Range of accessible WIMP speeds for each of the three mock experiments: XENON1T-like (solid blue), SuperCDMS-like (dashed green) and WArP-like (dot-dashed red). Each pair of lines corresponds to the maximum and minimum accessible WIMP speeds for a given experiment. The outermost dotted red lines show the accessible speeds for the adjusted parametrisation range described in Sec.\ \ref{sec:MomentumMethod2}.}
  \label{fig:Speeds}
\end{figure}

Figure \ref{fig:Speeds} shows the minimum and maximum accessible WIMP speeds for each experiment. All three experiments rapidly become insensitive to WIMPs with speeds less than \(\sim 1000 \kms\) as the WIMP mass drops below \(m_\chi \sim 10 \textrm{ GeV}\). According to the RAVE survey \cite{DMA19}, the local galactic escape speed lies in the interval \(v_\textrm{esc} \in \left[ 498, 608\right] \textrm{ km s}^{-1}\). The fastest WIMPs impinging on an experiment will then have speed \(v = v_{\textrm{lag}} + v_\textrm{esc}\). This is consistent with results from N-body simulations which indicate local escape velocities (in the Earth frame) of around \(800 \kms\) \cite{DMA43}. This suggests that the experiments considered here generically have a low sensitivity to such light WIMPs, producing too few events for accurate parameter reconstruction.

 We therefore use benchmark masses of \(50 \textrm{ GeV}\) and \(100 \textrm{ GeV}\). For WIMP masses heavier than this, \(v_\textrm{min}\) becomes roughly independent of \(m_\chi\), leading to large degeneracy in the \(m_\chi - \sigma_p\) plane \cite{DMDD50,DMDD51,DMDD97}. This will mean that for very heavy WIMPs, it will only be possible to place lower bounds on the WIMP mass. In fact, at the statistical precision presented here this is also true for most mock datasets generated with \(m_\chi = 100 \textrm{ GeV}\). As a benchmark cross-section, we use the value \(\sigma_p = 10^{-44} \textrm{ cm}^2\), which is still a factor of several below current exclusion limits. We assume that the local dark matter density is known exactly and use the value \(\rho_0 = 0.3 \textrm{ GeV cm}^{-3}\). As will be explained in Sec.\ \ref{sec:MomentumMethod2}, the precise values of \(\sigma_p\) and \(\rho_0\) are not particularly important due to the degeneracy between these two parameters. The total number of events from all three detectors combined typically ranges from around 300 to 600 for the benchmark parameters considered here.

In order to ensure the robustness of the method, we use three benchmark models for the velocity distribution:

\begin{enumerate}[(i)]
\item the Standard Halo Model (SHM), with \(\sigma = 156 \kms\) and \(v_\textrm{lag} = 230 \kms\);
\item a 50\% Standard Halo Model with a 50\% contribution from a dark disk (DD);
\item rescaled Via Lactea II data (VL-2).
\end{enumerate}

We model the dark disk velocity distribution as a Maxwellian with \(\sigma = 50 \kms\) and \(v_\textrm{lag} = 60 \kms\), similar to the typical values obtained by Ref.\ \cite{DMA20}. A 50\% contribution from the dark disk is at the upper limit of the range presented by Ref.\ \cite{DMA55} and we consider this as an extreme case. The third benchmark is the distribution function as extracted from the Via Lactea 2 (VL-2) N-body simulation \cite{VL2} and presented in Ref.\ \cite{DMA43}. It is averaged over galactic radius in the range \(7.5 < R < 9.5 \textrm{ kpc}\) and measured in bins of width \(10 \textrm{ m s}^{-1}\). VL-2 is a DM-only simulation and thus leads to a lower peak speed than the SHM. Including the effects of baryons should deepen the galactic potential and raise this peak speed closer to that observed in the Milky Way. In order for a fairer comparison, we therefore rescale the VL-2 data such that \(f_3(v)\) peaks at the same speed as in the SHM, allowing us to probe the departures from Maxwellian form which appear in N-body simulations.

\begin{figure}[t]
\centering
\includegraphics[width=0.5\textwidth]{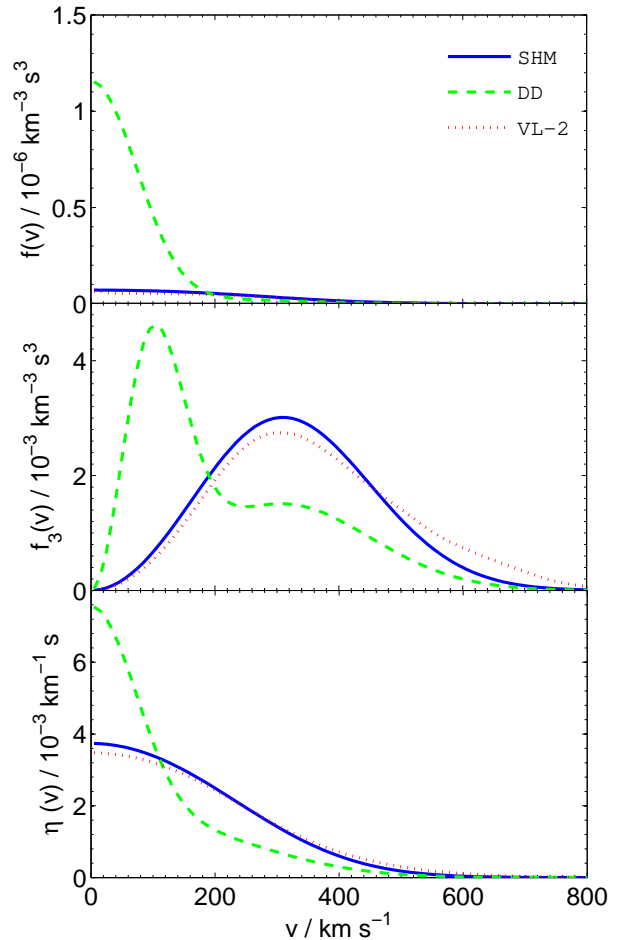}
\caption{1-D and 3-D speed distributions, \(f(v)\) and \(f_3(v)\), and mean inverse speed, \(\eta(v)\), for the 3 benchmark speed models with parameter values as given in Sec.\ \ref{sec:ParamBenchmarks} : Standard Halo Model (SHM - solid blue), Standard Halo Model + Dark Disk (DD - dashed green) and Via Lactea 2 (VL-2 - dotted red).}
  \label{fig:benchmarkf}
\end{figure}

These benchmark velocity distributions are illustrated in Fig. \ref{fig:benchmarkf}. The VL-2 data has the flattest velocity distribution with a tail extending beyond \(800 \kms\). This leads to a flatter spectrum and a larger number of events at higher energies than for the other two benchmark models. The SHM distribution produces roughly the same number of events as the VL-2 distribution, but with fewer events at high energy. In the dark disk model, however, the value of \(v_\textrm{lag}\) is much smaller. This means that WIMPs typically have much lower speeds and many have insufficient energy to overcome the thresholds of the detector. This results in fewer observed events and a steeper recoil spectrum.

\subsection{Parameter Exploration}
\label{sec:ParamExplore}
We generate mock datasets for each set of benchmark WIMP parameters and then analyse these using Markov Chain Monte Carlo (MCMC) techniques. For a given set of physical parameters, \(\theta_\textrm{true} = \{m_\chi, \sigma_p, f(v)\}\), we calculate the expected number of events for each detector, \(N_\textrm{e}\), to an accuracy of 0.1 events:

\begin{equation}
N_\textrm{e} = m_\textrm{det}  t_\textrm{exp} \int_{E_\textrm{min}}^{E_\textrm{max}} \frac{\textrm{d}R}{\textrm{d}E}(E) \, \textrm{d}E.
\end{equation}

For each experiment, the number of events actually observed, \(N_\textrm{o}\), is drawn from a Poisson distribution with mean \(N_\textrm{e}\). Energies for the \(N_\textrm{o}\) events are then drawn from the normalised differential event rate:

\begin{equation}
P(E) = \frac{m_\textrm{det} t_\textrm{exp}}{N_\textrm{e}} \, \frac{\textrm{d}R}{\textrm{d}E}(E).
\end{equation}

We then use the publicly available \textsc{CosmoMC} code \cite{MCMC2} to generate Markov chains to explore the posterior distribution of the parameter space. This parameter space typically comprises \(m_\chi\), \(\sigma_p\) and additional parameters associated with the speed or momentum distribution of the WIMPs. The likelihood function used to generate the Markov chains is the same form used by CDMS \cite{DMDD91} and XENON100 \cite{DMDD90}, which for a single experiment is:

\begin{equation}
\mathcal{L}_1 = \frac{N_\textrm{e}^{N_\textrm{o}}\, \textrm{e}^{-N_\textrm{e}}}{N_\textrm{o}!} \, \prod_j^{N_\textrm{o}} P(E_j).
\end{equation} 
The full likelihood, \(\mathcal{L}\), is then the product of the likelihoods for the three separate experiments.

For each set of benchmark parameters, we generate 250 mock datasets. For each realisation, we perform \(3 \times 10^5\) Markov chain steps using the Metropolis-Hasting algorithm. Each chain is then thinned by a factor of 50 and the initial \(10^4\) chain positions removed as burn-in.  The results obtained were robust to doubling of either the thinning factor or the number of positions taken as burn-in. For a well-mixed chain, the distribution of chain positions as a function of the parameters, \(n(\theta)\), should be proportional to the local likelihood: \(n(\theta) \propto \mathcal{L}(\theta)\). For flat priors, this is then proportional to the posterior probability of the parameters \(\theta\). However, in order to ensure that the posterior is well-explored (particularly in cases where it may be multimodal), we perform the MCMC at a temperature \(T = 2\), using the high temperature likelihood function \(\mathcal{L}_\textrm{T} = \mathcal{L}^{\nicefrac{1}{\textrm{T}}}\) \cite{MCMC8}. The distribution of chain positions obtained at the higher temperature is then \(n_\textrm{T}(\theta) \propto \mathcal{L}^{\nicefrac{1}{T}}(\theta)\). We recover the distribution of positions at \(T=1\) by `cooling' the chain:

\begin{equation}
n(\theta) = n_\textrm{T}(\theta) \mathcal{L}^{1 - \nicefrac{1}{\textrm{T}}}(\theta).
\end{equation}

In order to obtain parameter limits from the mock datasets, we construct minimal credible intervals for the parameters of interest \cite{MCMC5}. In order to construct a \(p\%\) confidence interval, we find a value of the marginalised likelihood for which \(p\%\) of chain samples exceed this value. All parameter values for which the marginalised likelihood exceeds this limit are included in the confidence interval. This allows us to investigate the coverage properties of different methods and parameters. For each mock dataset, we record whether or not the true parameter value lies within this confidence interval. If a method produces consistent error estimates, the true parameter value should lie within the \(p\%\) confidence interval for \(p\%\) of realisations. This is referred to as exact coverage.

We note that MCMC is typically poor at finding the best-fit point in parameter space \cite{MCMC2}, so in order to obtain an estimate of parameters, we use the mean likelihood, as described in Ref.\ \cite{MCMC2}. We bin the parameter of interest and average the likelihood of all points within each bin. This mean likelihood is then smoothed on the width of the bins and the parameter value which maximises the mean likelihood is taken as a best estimate of the underlying parameter.

As an estimate of an error on a parameter, we calculate its standard deviation within a chain, \(\sigma\). While this does not correspond to the Gaussian error on the parameter (particularly in the case of multimodal and skewed posterior distributions), it can be used as a rough estimate of the uncertainty of a particular reconstruction. We can also use this to calculate the `pull' statistic \(\Delta\) (see e.g.\ Ref.\ \cite{Pulls}),

\begin{equation}
\Delta = \frac{\theta_\textrm{rec} - \theta_\textrm{true}}{\sigma} \,,
\end{equation}   
where \(\theta_\textrm{rec}\) and \(\theta_\textrm{true}\) are the reconstructed and true values of the parameter of interest respectively. The \(\Delta\) statistic quantifies the statistical deviation of the reconstruction from the true value. For a large number of mock datasets, the distribution of \(\Delta\) should have a mean of zero and a standard deviation of unity.

We sample the WIMP mass and cross-section logarithmically in the ranges \([10, 1000] \textrm{ GeV}\) and \([10, 10000] \times 10^{-47} \textrm{ cm}^2\) respectively, with log-flat priors on both. Unless otherwise stated, parameters associated with the WIMP distribution are sampled linearly between zero and their maximum possible value allowed by normalisation, with linearly-flat priors.

\section{Speed Parametrisation}
\label{sec:SpeedMethod}
We follow the method of Peter \cite{DMDD3} and parametrise the WIMP speed distribution (in the Earth frame), as a series of \(N\) bins of constant value, with bin edges \(\{ \tilde{v}_i\}\):

\begin{equation}
f(v) = \sum_{i = 1}^N g_i \, W(v;\tilde{v}_i,\Delta v) \,,
\end{equation}
where the top-hat function, W, is defined as:

\begin{equation}
W(v;\tilde{v}_i,\Delta v) =
\begin{cases}
   1 &  v \in [\tilde{v}_i,\tilde{v}_i+\Delta v] \\
   0  & \text{otherwise}
  \end{cases} \,.
\end{equation}
Here, we consider \(N = 5\) bins, and take the bins to be of constant width, \(\Delta v = 200 \kms\), covering the range \(v = 0 - 1000 \kms\). Imposing normalisation of the speed distribution, we obtain the following constraint on the \(\{g_i\}\):
\begin{equation}
\label{eq:Normg}
\sum_{i = 1}^N g_i \, \left[(\tilde{v}_i + \Delta v)^3 - \tilde{v}_i^3\right]/3 = 1 \,.
\end{equation}
For notational convenience, we also define

\begin{equation}
\hat{g}_i = g_i \, \left[(\tilde{v}_i + \Delta v)^3 - \tilde{v}_i^3\right]/3 \,,
\end{equation}
such that the normalisation condition becomes
\begin{equation}
\label{eq:ghat}
\sum_{i = 1}^N \hat{g}_i = 1.
\end{equation}

\subsection{Results}

\begin{figure}[t]
\centering
  \includegraphics[width=0.5\textwidth]{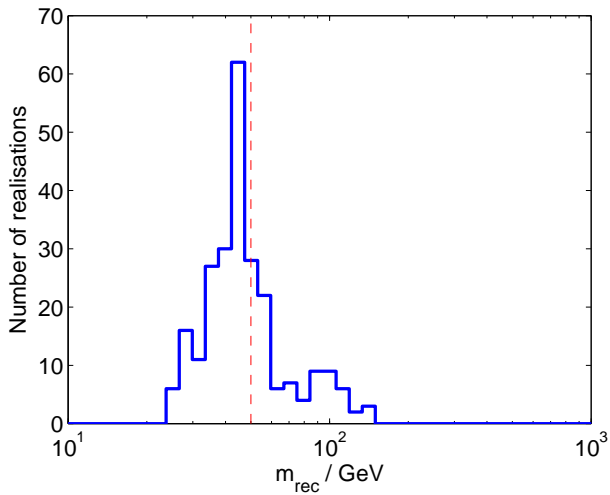}
  \caption{WIMP masses reconstructed using the speed parametrisation method from 250 realisations. The benchmark speed distribution is the SHM. The true mass of 50 GeV is shown as a dashed vertical line.} 
  \label{fig:Speed_MassFit}
\end{figure}

For comparison with later methods, we consider here a single benchmark model: \(m_\chi = 50 \textrm{ GeV}\) and the SHM. Fig.\ \ref{fig:Speed_MassFit} shows the fitted values for the WIMP mass, \(m_\textrm{rec}\), obtained from 250 mock datasets. This distribution shows a peak around 45 GeV, as well as a significant number of datasets reconstructed at \(\sim 100 \textrm{ GeV}\). As pointed out by Ref.\ \cite{DMDD40}, some mock datasets will not be representative of the underlying benchmark parameters, having more events at high energies than expected, for example. This can lead to `bad' reconstructions with a fitted WIMP mass higher than the benchmark value. Thus, the reconstructions near \(100 \textrm{ GeV}\) do not necessarily signify a failure of the reconstruction method.

\begin{figure}[t]
\centering
  \includegraphics[width=0.5\textwidth]{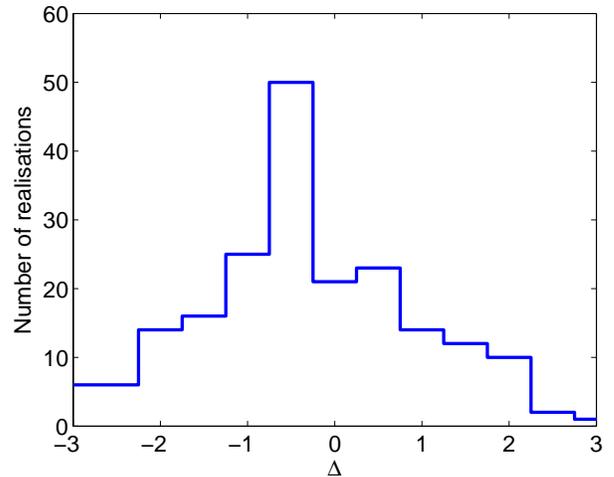}
  \caption{Distribution of the \(\Delta\) statistic, defined in Sec.\ \ref{sec:ParamExplore}, for 250 realisations using the speed parametrisation method. The benchmark speed distribution is the SHM, with a 50 GeV WIMP.} 
  \label{fig:Speed_Delta}
\end{figure}

However, a coverage study of 68\% and 95\% confidence intervals for this method shows significant under-coverage: \(36 \pm 3 \%\) coverage and \(63 \pm 3 \%\) coverage respectively for the two intervals. This indicates that while the mass reconstructions appear to be distributed close to the true value, the corresponding error estimates must be too small. Fig.\ \ref{fig:Speed_Delta} shows the \(\Delta\) distribution for the 250 realisations of this method. The standard deviation of \(\Delta\) is \(\sigma_\Delta = 3.5\), which is significantly greater than unity, showing that the speed parametrisation method significantly underestimates the errors on reconstructed values. As demonstrated in Ref.\ \cite{DMDD3}, the problem of poor reconstructions using this method does not appear to be significantly improved by increasing the number of bins in the speed parametrisation and worsens for more complicated speed distributions.

\begin{figure}[t]
\centering
  \includegraphics[width=0.5\textwidth]{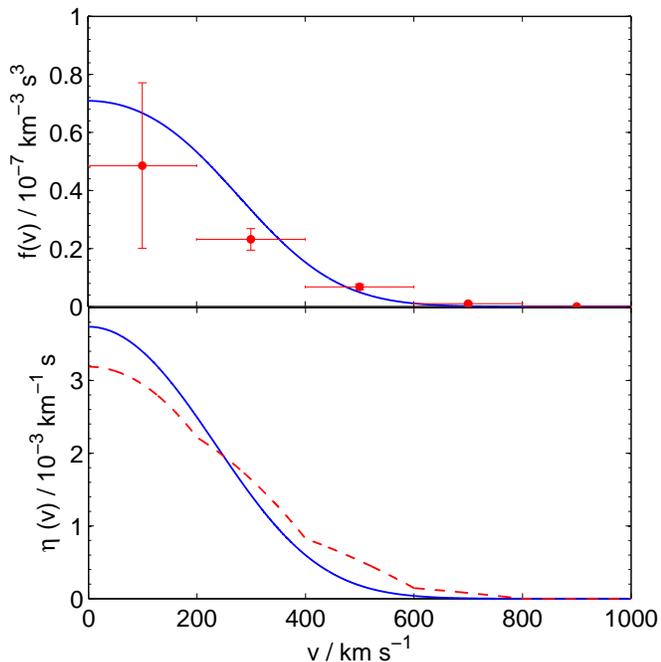}
\caption{Reconstructed speed distribution, \(f(v)\), and mean inverse speed, \(\eta(v)\), using the speed parametrisation method. The benchmark model used was a 50 GeV WIMP with an SHM speed distribution. The upper pane shows the underlying SHM speed distribution (solid blue) and the fitted values of the speed bin parameters (red points). The errors on the bin values are within-chain standard deviations as described in Sec.\ \ref{sec:ParamExplore}. The lower pane shows the mean inverse speed corresponding to these fitted values (dashed red line) and the true mean inverse speed (solid blue). }
  \label{fig:Speed_Recon}
\end{figure}

Fig.\ \ref{fig:Speed_Recon} shows the reconstructed speed distribution for a typical realisation using this method. The reconstructed mass value is \(\log_{10} (m_\textrm{rec} / \textrm{ GeV}) = 1.48 \pm 0.06\), compared to the benchmark value \(\log_{10} (m_\chi / \textrm{ GeV}) = 1.699\). The mean inverse speed is under-estimated in the range \(0 - 200 \kms\) and slightly over-estimated at higher speeds. However, the reduced \(m_\textrm{rec}\) increases the minimum accessible speed of the experiments, meaning that the experiments are less sensitive to the shape of the speed distribution at low speeds. Moreover, a reduced value of the reconstructed mass serves to steepen the spectrum, reconciling the flattened \(\eta(v_\textrm{min})\) at high speeds with the data. This is because varying the mass of the WIMP `rescales' the spectrum, due to the relation \(E_\textrm{R} \propto \mu_{\chi N} v_{\textrm{min}}^2\).

 \begin{figure}[t]
\centering
  \includegraphics[width=0.5\textwidth]{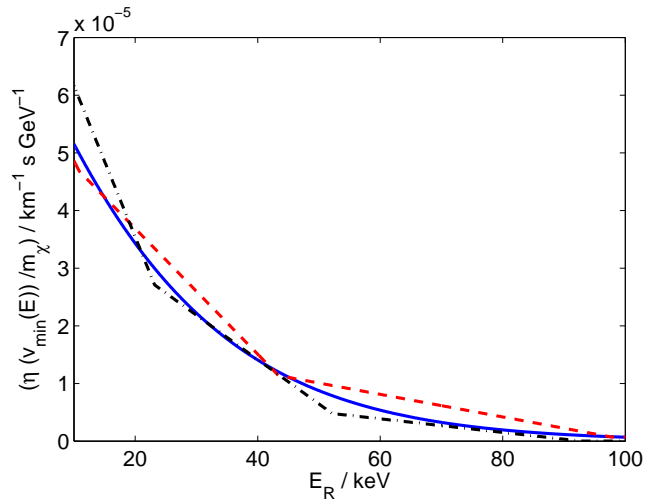}
\caption{The rescaled mean inverse speed, \(\eta/m_\chi\), measured in the SuperCDMS-like experiment as a function of recoil energy, \(E_\textrm{R}\). The same mock dataset was used as in Fig.\ \ref{fig:Speed_Recon}. The underlying Standard Halo Model distribution (solid blue) uses the true WIMP mass of 50 GeV, as does the binned approximation to the SHM (dashed red). The reconstructed mean inverse speed (dot-dashed black) uses the reconstructed value of 30 GeV.}
  \label{fig:Compare}
\end{figure}

In Fig.\ \ref{fig:Compare}, we plot \(\eta/m_\chi\) as a function of recoil energy, \(E_\textrm{R}\), for the SuperCDMS-like experiment. We rescale \(\eta\) by \(1/m_\chi\) because this factor appears in the event rate and we are then able to compare the spectra of events from different models. The solid line shows the mean inverse speed in the SHM, using the true WIMP mass of 50 GeV. We also show a binned approximation to the SHM (dashed line) obtained using the true values of the bin parameters \(\{g_i\}\) and the true WIMP mass. Finally, we show the reconstructed mean inverse speed (dot-dashed line) using the reconstructed WIMP mass of 30 GeV. We see that the binned approximation to the SHM, which should represent a `good' reconstruction, actually recovers the spectrum poorly compared to the reconstructed values. In particular, we note the energy range of the experiment spans two bins in the binned approximation to the SHM, but three bins in the MCMC reconstruction, allowing a closer approximation to the true spectrum.

Thus, the speed distribution parameters can be explored to provide a good fit to the data, with the reconstructed mass varying to compensate. As can be seen from Fig.\ \ref{fig:Compare}, for a fixed bin width in velocity space, the size of bins in energy space can be reduced by moving to lower masses. This should allow a closer fit to the data and may explain why there appears to be a bias towards lower mass values.

\section{Momentum parametrisation for a single experiment}
\label{sec:MomentumMethod1}

When considering the speed distribution of the WIMPs, we see that each experiment has a different range of sensitivity and that varying the WIMP mass changes this range. However, we can instead consider a `reduced WIMP-nucleus momentum,'
\begin{equation}
\textbf{p}_N = \mu_{\chi N} \textbf{v} \,,
\end{equation}
defined separately for each target nucleus. We now note that the accessible range in \(\textbf{p}_N\) for each experiment is independent of the WIMP mass:
\begin{equation}
p_\textrm{min}(E_R) = \mu_{\chi N} v_\textrm{min}(E_R) = \sqrt{\frac{m_N E_R}{2}} \,.
\end{equation}

We therefore rewrite the differential event rate in terms of the new momentum variable:
\begin{equation}
\frac{\textrm{d}R}{\textrm{d}E_R} = \frac{\rho_0 \sigma_p \mu_{\chi N}}{2 \mu_{\chi p}^2 m_\chi} A^2 F^2(E_R) \tilde{\eta}(p_{\textrm{min}})\,,
\end{equation}
where \(\tilde{\eta}\) is the mean inverse \textit{momentum} associated with the reduced momentum distribution, \(\tilde{f}(\textbf{p})\):
\begin{equation}
\tilde{\eta}(p_{\textrm{min}}) = \int_{p_{\textrm{min}}}^\infty \frac{\tilde{f}(\textbf{p})}{p}\, \textrm{d}^3\textbf{p} = \frac{1}{\mu_{\chi N}}\eta(p_\textrm{min}/\mu_{\chi N}).
\end{equation}
The event rate can be rewritten as:
\begin{equation}
\frac{\textrm{d}R}{\textrm{d}E_R} = \frac{\rho_0}{2} D(\sigma_p,m_\chi,m_N) A^2 F^2(E_R) \tilde{\eta}(p_{\textrm{min}}) \,,
\end{equation}
where we have defined
\begin{equation}
D(\sigma_p,m_\chi,m_N) = \frac{\sigma_p \mu_{\chi N}}{\mu_{\chi p}^2 m_\chi} \,,
\end{equation}
which encodes all information about the WIMP mass and cross-section and controls the overall scale of the event rate. 

We can again define a directionally averaged momentum distribution, \(\tilde{f}(p) = f(p/\mu_{\chi N})/\mu_{\chi N}^3\), and parametrise this in terms of 5 constant bins, with bin values \(\{h_i\}\). We parametrise \(\tilde{f}(p)\) only over the range of sensitivity of the experiment: \(p \in \left[p_a, p_b\right]\), where \(p_{a,b} = p_\textrm{min}(E_\textrm{min,max})\). This means that we need not make any assumptions about the distribution function outside the range of sensitivity of the experiment. However, we still wish to impose some normalisation constraint on the momentum distribution parameters. Each experiment now probes a well-defined (but unknown) fraction of WIMPs, \(\alpha_N\), given by

\begin{equation}
\alpha_N = \int_{p_a}^{p_b} f(p) \, p^2 \textrm{d}p\,.
\end{equation}
The momentum parameters are therefore normalised according to
\begin{equation}
\sum_{i = 1}^N \hat{h}_i = \alpha_N \,,
\end{equation}
where \(\hat{h}_i\) is defined analogously to \(\hat{g}_i\) in Eq.\ \ref{eq:ghat}. We absorb the unknown \(\alpha_N\) into \(D\), such that the momentum distribution parameters, \(\{\hat{h}_i/\alpha_N\}\), are normalised to unity and

\begin{equation}
\label{eq:D}
D(\sigma_p,m_\chi,m_N) = \alpha_N \frac{\sigma_p \mu_{\chi N}}{\mu_{\chi p}^2 m_\chi}\,.
\end{equation}

Finally, it is necessary to introduce a parameter \(A\) which models the constant contribution to \(\eta\) from WIMPs with momenta greater than \(p_b\):

\begin{equation}
A = \int_{p_\textrm{min}(E_\textrm{max})}^\infty \frac{\tilde{f}(\textbf{p})}{p}\, \textrm{d}^3\textbf{p}\,.
\end{equation}
Because the precise form of \(\tilde{f}(p)\) above the upper energy threshold is undetermined by the experiment, the contribution of \(A\) to the normalisation, \(\alpha_N\), cannot be calculated and is therefore not considered. Instead, we include conservative constraints on \(A\) such that its contribution alone cannot exceed the normalisation of \(\tilde{f}(p)\):
\begin{equation}
A < (p_\textrm{min}(E_\textrm{max}))^{-1}.
\end{equation}
We also note that
\begin{equation}
\int_{p_a}^{p_b} \frac{\tilde{f}(\textbf{p})}{p} \, \textrm{d}^3\textbf{p} \leq \frac{\alpha_N}{p_b} \,,
\end{equation}
and thus impose the following additional constraint on the paramaters:
\begin{equation}
\frac{1}{\alpha_N}\left[\eta(p_a) - \eta(p_b)\right] \leq \frac{1}{p_b}\,.
\end{equation}

We therefore perform parameter reconstructions using the parameters \(D\), \(\{h_i/\alpha_N\}\) and \(A/\alpha_N\). Because the fraction of high momentum WIMPs is expected to be relatively low, we sample the parameter \(A\) logarithmically, with a log-flat prior.

\subsection{Results}

\begin{figure}[t]
\centering
  \includegraphics[width=0.5\textwidth]{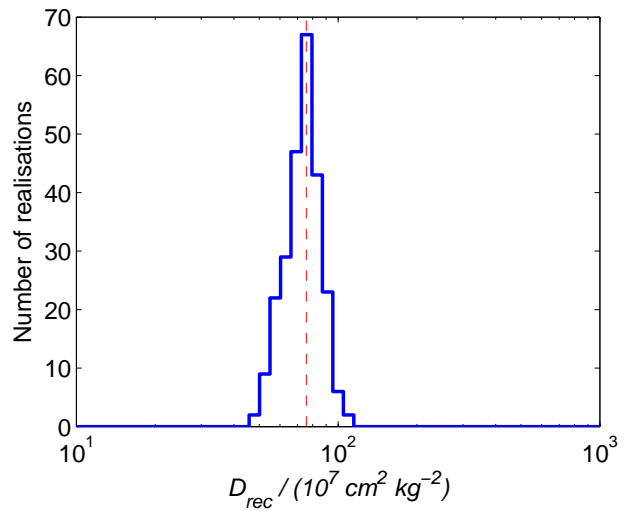}
  \caption{Reconstructed values for the scale parameter, \(D_\textrm{rec}\), for the Argon experiment using the momentum parametrisation method from 250 realisations. The benchmark speed distribution is the SHM. The value of \(D_\textrm{true} = 75.6 \times 10^7 \textrm{ cm}^2 \textrm{ kg}^{-2}\) is shown as a dashed vertical line.}
   \label{fig:Argon1}
\end{figure}

We consider again a single set of benchmark parameters, namely a \(50 \textrm{ GeV}\) WIMP with an SHM speed distribution. We apply the momentum parametrisation to mock datasets from the WArP-like Argon experiment. The reconstructed \(D\) values, \(D_\textrm{rec}\),  are shown in Fig.\ \ref{fig:Argon1} in units of \(10^7 \textrm{ cm}^2 \textrm{ kg}^{-2}\). In all reconstructions, the posterior distribution is unimodal, having separate parameters to describe the scale (\(D\)) and shape (\(\{h_i\}\)) of the event rate. The number of reconstructions is peaked at the correct value, however, the distribution does not appear to be symmetric. In fact, the average reconstructed value is \(\log_{10}(D)_\textrm{rec} = 1.865 \pm 0.004\), compared to the input value of \(\log_{10}(D)_\textrm{true} = 1.878\). This represents a slight bias (of less than \(1\%\)) towards smaller values of \(\log_{10}(D)\).

However, this is smaller than the typical statistical uncertainty in a single reconstruction, which is \(\sim 4\%\). In addition, this method results in \textit{over}coverage of the true parameter, with values of \(76 \pm 2 \%\) and \(98 \pm 1\%\) respectively for the \(68\%\) and \(95\%\) confidence intervals. This method therefore allows us to place reliable conservative estimates on the parameter \(D\).

\begin{figure}[t]
\centering
\includegraphics[width=0.5\textwidth]{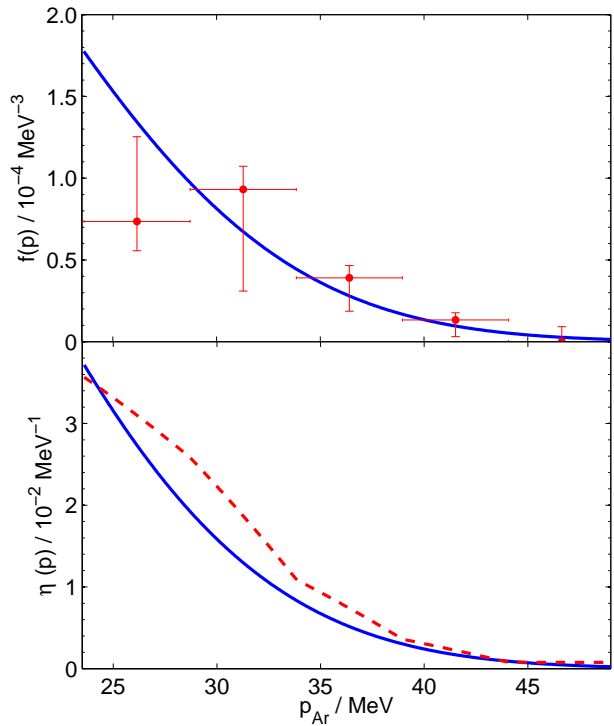}
\caption{Reconstructed momentum distribution for a single Argon experiment using a benchmark of a 50 GeV WIMP and the SHM.  The upper pane shows the SHM momentum distribution (solid blue) and reconstructed bin values (red points). Because the posterior is unimodal, we also display vertical errorbars showing the extent of the 68\% confidence region for each bin. Note that these errors are strongly correlated. The lower pane shows the corresponding reconstructed mean inverse momentum (dashed red) and the mean inverse momentum in the SHM (solid blue). The underlying distribution has been rescaled by \(1/\alpha_{\textrm{Ar}}\) for comparison to the reconstructed values.}
  \label{fig:Ar_Momentum}
\end{figure}

We show in Fig.\ \ref{fig:Ar_Momentum} the reconstructed momentum distribution and mean inverse momentum for a typical realisation, for which the reconstructed \(D\) value is \(1.81_{-0.05}^{+0.09}\). The underlying momentum distribution has been rescaled by \(1/\alpha_\textrm{Ar}\) to allow a comparison to the reconstructed values. We see that the the momentum distribution is well reconstructed and the mean inverse momentum is accurately recovered at low and high momenta. In the middle of the momentum range, however, \(\tilde{\eta}(p_\textrm{min})\) exceeds the true value. Because only a single experiment is being used, the measured spectrum is particularly susceptible to Poisson fluctuations. The mock dataset used here has a slight excess of events around \(E_\textrm{R} \approx 60 \textrm{ keV}\), corresponding to \(p_\textrm{Ar} \approx 30 \textrm{ MeV}\), which may explain the reconstructed excess. 

In addition, this may be a consequence of the particular parametrisation. The constant-bin parametrisation of \(\tilde{f}(p)\) leads to a parametrised \(\tilde{\eta}(p_\textrm{min})\) which is concave downwards in each bin, while the underlying function is strictly convex downwards in this region. Thus, \(\tilde{\eta}(p_\textrm{min})\) tends to be slightly overestimated, leading the scale parameter \(D\) to be reduced to compensate for this. With datasets containing more events, the number of bins could be increased, in order to reduce this bias on \(D\) and maintain it at below the level of the statistical uncertainty.

\section{Momentum parametrisation for several experiments}
\label{sec:MomentumMethod2}

The reduced momentum method allows us to extract information from a single experiment, making no assumptions about the underlying velocity (or momentum) distribution. However, information about the mass and cross-section are encoded in the parameter, \(D\), and cannot be extracted using a single experiment alone.

Because a different momentum variable \(p_N\) can be defined for each experiment, it is necessary to choose a single experiment and parametrise the momentum distribution defined with respect to that experiment. It may be necessary to adjust the lower and upper limits of the parametrisation (beyond the values of \(E_\textrm{min}\) and \(E_\textrm{max}\) used in the experiment) to accommodate as much of the data as possible from all experiments. In the single experiment considered in Section \ref{sec:MomentumMethod1}, the WIMP-Ar momentum was parametrised in the range \(p_\textrm{Ar} \in \left[23.6, 49.2\right] \textrm{ MeV}\), to match the sensitivity of the Argon experiment. However, as can be seen in Fig.\ \ref{fig:Speeds}, this sensitivity window does not match that of the other experiments. If we extend this interval, and parametrise in the range \(p_\textrm{Ar} \in \left[3.6, 53.0\right] \textrm{ MeV}\), we can enclose the sensitivity regions of all three experiments as closely as possible, as shown by the dotted curves in Fig.\ \ref{fig:Speeds}. We again use 5 bins in momentum space, with an additional parameter to control a constant offset.

In theory, any of the three experiments could have been chosen to define the momentum variable. However, some choices of experiment are less practical. For example, in order to use the XENON1T-like experiment, it would be necessary to parametrise the momentum over the range \(p_\textrm{Xe} \in \left[11 , 162\right] \textrm{ MeV}\). This is because at high WIMP masses the remaining two experiments have maximum accessible speeds of \(\sim 500 \kms\). This corresponds to very high values of the WIMP-Xe reduced momentum because of Xenon's comparatively higher mass. A large number of bins would be required to cover this wide momentum range and accurately model structures in the distribution function. Owing to the Galactic escape speed, many of these bins would have a value of zero, making parametrisation with respect to the XENON1T-like experiment a poor choice.

By comparison, using the WArP-like Argon experiment allows us to parametrise only as much of the momentum space as required to accommodate data from all three experiments. In the speed parametrisation method, the WIMP mass could be varied to adjust the range of speeds accessible to the experiments. This reduces the sensitivity of the likelihood to some of the speed bin parameters, meaning that these parameters can be adjusted with little effect on the likelihood value. This results in a spurious freedom in the remaining bin parameters, which can be varied to achieve a good fit to the data. As demonstrated in Sec.\ \ref{sec:SpeedMethod}, this results in a bias towards lower values of the reconstructed WIMP mass in order to reduce the size of the bins in energy space. Using the momentum parametrisation method, we reduce this effect by ensuring that the likelihood is sensitive to all momentum bin parameters as much as possible, as the accessible speed range of the analysis tracks more closely the accessible speed range of the experiments. In addition, for a fixed bin width in momentum space, the bin width in energy space is much less sensitive to the WIMP mass.

Unfortunately, this method does not allow the WIMP-nucleon cross-section to be extracted; because the contributing WIMP fraction, \(\alpha\), is unknown, we can only obtain a lower bound. This is a fundamental limitation of any method which makes no assumptions about the underlying speed distribution. Without knowing the fraction of WIMPs with speeds within the signal window of the experiment, we cannot determine the cross-section. However, the cross-section appears in the event rate only through the degenerate combination \(\sigma_p \rho_0\). Estimates of \(\rho_0\) from mass modelling of the Milky Way and from stellar kinematics tend to have a large uncertainty, with \(\rho_0\) typically lying in the range \(0.2 - 0.4 \textrm{ GeV cm}^{-3}\) (e.g.\ Ref.\ \cite{DMA13,DMA36}). More recent analyses have found somewhat larger values, though still with large uncertainties. Salucci et al.\ \cite{DMA34} find \(\rho_0 = 0.430 \pm 0.209 \textrm{ GeV cm}^{-3}\), while Garbari et al.\ \cite{DMA48} obtain the value \(\rho_0 = 0.85_{-0.50}^{+0.57} \textrm{ GeV cm}^{-3}\). Thus, any estimate of the WIMP-nucleon interaction cross-section would have an inherent uncertainty in any case.

\subsection{Results}

\begin{figure}[t]
\centering
  \includegraphics[width=0.5\textwidth]{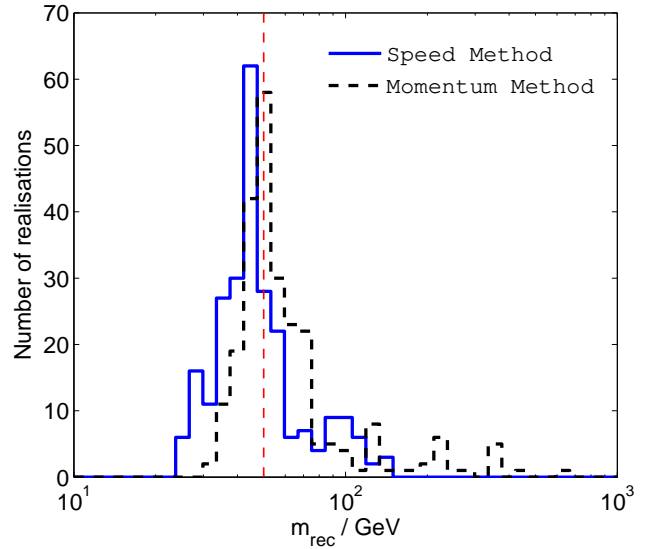}
  \caption{WIMP masses reconstructed using the speed and momentum parametrisation methods from 250 realisation. The benchmark speed distribution is the SHM. The true mass of 50 GeV is shown as a dashed vertical line.} 
  \label{fig:both}
\end{figure}

We first compare results for the momentum method to the speed parametrisation method presented in Section \ref{sec:SpeedMethod}. We use the same mock datasets generated for the \(50 \textrm{ GeV}\), SHM benchmark presented previously. The results of both the momentum and speed methods are shown in Fig.\ \ref{fig:both}. In the case of the momentum method, the distribution of realisations is now more closely peaked around the true mass of \(50 \textrm{ GeV}\). Furthermore, the momentum method produces substantially improved coverage properties, as summarised in Table \ref{tab:CoverageComparison}. It should be noted that compared to the speed method, the momentum method leads to a larger number of reconstructions at high WIMP mass. It is not clear whether this signals a failure of the momentum method in certain cases or whether these are representative of `bad' reconstructions, as will be discussed shortly. 

\begin{figure}[t]
\centering
  \includegraphics[width=0.5\textwidth]{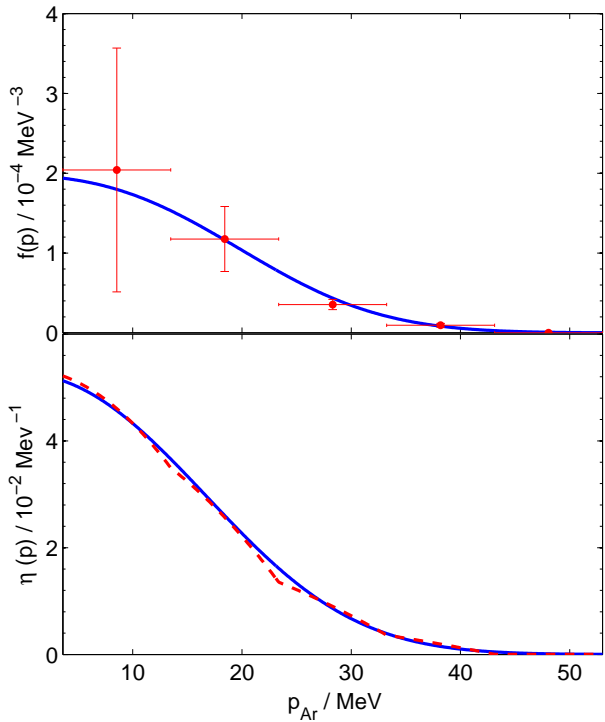}
\caption{Reconstructed momentum distribution from all three mock experiments using a benchmark of a 50 GeV WIMP and the SHM. The upper pane shows the SHM momentum distribution (solid blue) and reconstructed bin values (red points). The errors on the bin values are within-chain standard deviations as described in Sec.\ \ref{sec:ParamExplore}.  The lower pane shows the corresponding reconstructed mean inverse momentum (dashed red) and the mean inverse momentum in the SHM (solid blue). The reconstructed values have been rescaled by \(\alpha\) for comparison to the true distribution.}
  \label{fig:50SHM_momentum}
\end{figure}

Fig.\ \ref{fig:50SHM_momentum} shows the reconstructed WIMP-Argon momentum distribution using the same mock dataset as used for Fig.\ \ref{fig:Speed_Recon}. The benchmark distributions have been rescaled by \(\alpha\) so that they can be compared to the reconstructed values. In this case, \(\alpha = 0.995\), so we can reconstruct both the mass and cross section accurately: \(\log_{10} (m_\textrm{rec} / \textrm{ GeV}) = 1.62 \pm 0.31\) and \(\log_{10} (\sigma_p / 10^{-47} \textrm{ cm}^2) = 2.99 \pm 0.18\), compared to the true values of \(\log_{10} (m_\textrm{true} / \textrm{ GeV}) = 1.699\) and \(\log_{10} (\sigma_p / 10^{-47} \textrm{ cm}^2) = 3.0\). While there is no way to know \textit{a priori} whether \(\alpha\) will be close to unity, the accurate reconstruction of the mass, cross-section and momentum distribution show that momentum parametrisation can offer a significant improvement over the speed parametrisation method.

\begin{table}[t]
  \begin{center}
\begin{ruledtabular}
    \begin{tabular}{lll}
    
    & Speed Method & Momentum Method \\
    \hline
    68\% Coverage & \(36 \pm 3 \%\) & \(71 \pm 3 \%\) \\
    95\% Coverage & \(63 \pm 3\%\) & \(92 \pm 2 \%\) \\
    \end{tabular}
\end{ruledtabular}
  \end{center}
  \caption{Coverage statistics for the speed and momentum parametrisation methods for a 50 GeV SHM benchmark model.}
\label{tab:CoverageComparison}
\end{table}

\begin{figure}[t]
\centering
\includegraphics[width=0.5\textwidth]{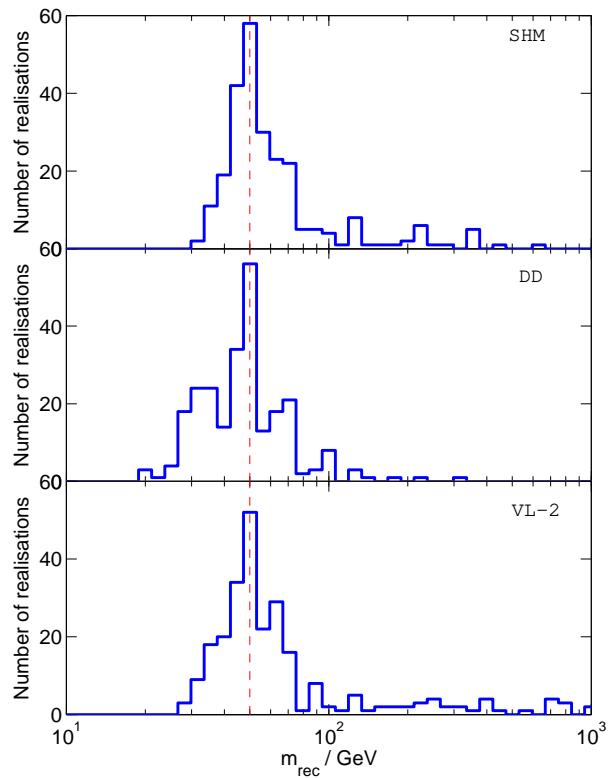}
\caption{Distribution of reconstructed masses, \(m_\textrm{rec}\), using the momentum method for 250 reconstructions. The true mass of 50 GeV is shown as a dashed vertical line.}
  \label{fig:recons50}
\end{figure}

\begin{figure}[t]
\centering
\includegraphics[width=0.5\textwidth]{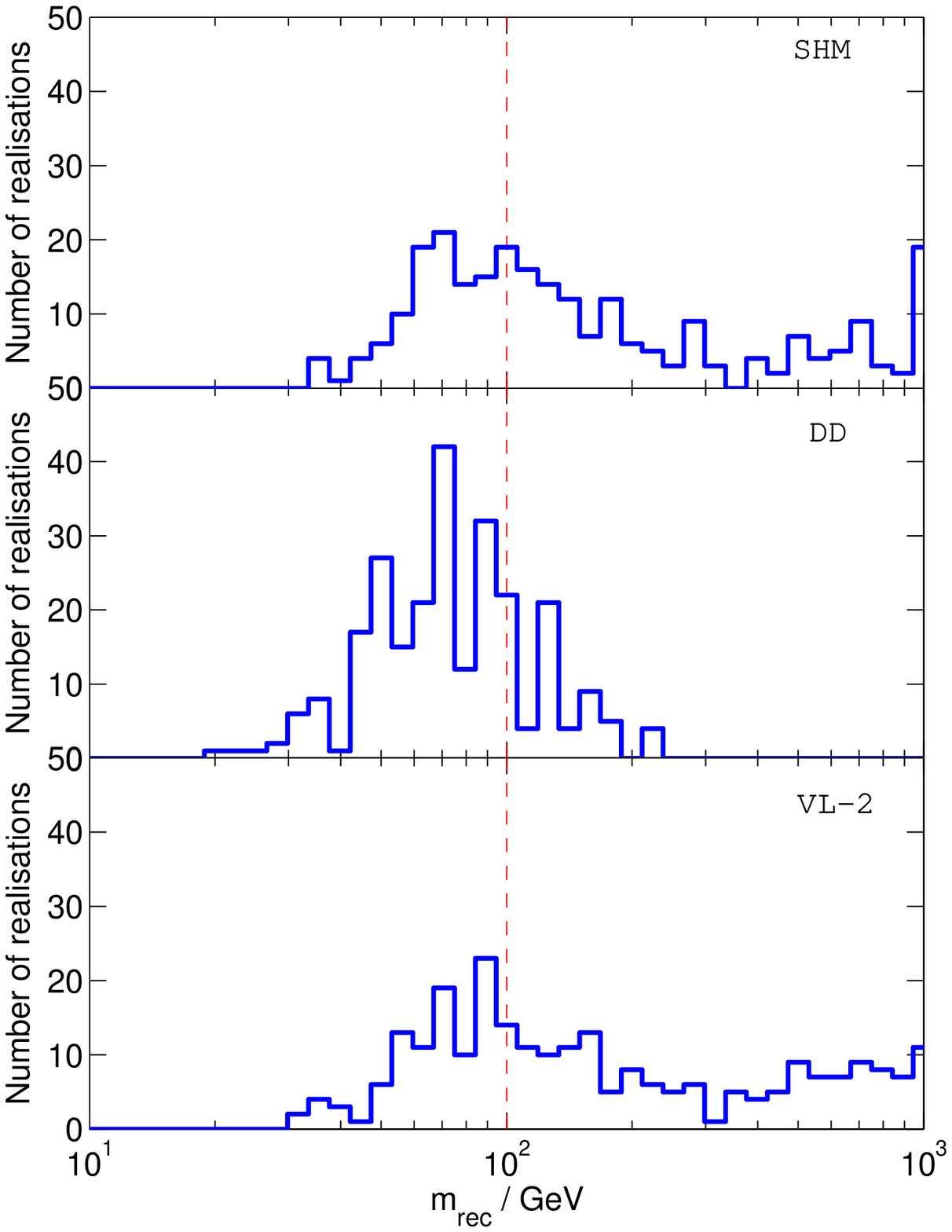}
\caption{As Fig.\ \ref{fig:recons50} for \(m_\chi = 100 \textrm{ GeV}\).}
  \label{fig:recons100}
\end{figure}

We now present the results of the momentum method for all six sets of benchmark parameters, 50 GeV and 100 GeV WIMPs, with Standard Halo Model, dark disk and Via Lactea-2 distribution functions. The distributions of reconstructed masses are shown in Fig.\ \ref{fig:recons50} for the 50 GeV WIMP and Fig.\ \ref{fig:recons100} for the 100 GeV WIMP.

For the 50 GeV benchmark, the distribution of reconstructions is peaked at the true value, though in all three cases there are a number of datasets reconstructed at higher masses. For some of the mock datasets, the posterior distribution for the WIMP mass is multimodal, with a peak near the true value as well as a peak above \(\sim 100 \textrm{ GeV}\). For reconstructions using a fixed speed (or momentum) distribution, these may correspond to `bad' reconstructions, as mentioned previously, in which the spectrum of events is flatter than expected. When the momentum distribution is allowed to vary, as here, the event rate can be well fit by more than one region of the mass parameter space.  We also note a larger number of reconstructions at high masses in the case of the VL-2 benchmark. This is because of the flatter recoil spectrum in this case, which is more easily mimicked by a higher WIMP mass.

For the 100 GeV benchmark, the SHM and VL-2 models show similar structures, with a broad peak of reconstructions at or near the correct values, as well as a smaller tail up to masses of 1000 GeV, the upper limit of the prior. The 100 GeV datasets contain fewer events than their 50 GeV counterparts, so we would expect the spread of reconstructed values to be broader. Also, as previously noted, as the WIMP mass exceeds the mass of the target nucleus, the shape of the event spectrum becomes roughly independent of the WIMP mass. The largest nuclear mass used here is \(A_\textrm{Xe} = 131\), meaning that for values of \(m_\textrm{rec}\) significantly above \(m_\chi \approx 131 \textrm{ amu} \approx 122 \textrm{ GeV}\), the posterior distribution becomes roughly flat. Reconstructions in the high-mass tail occur when the maximum of the posterior occurs in this approximately flat region, and we expect the tail to extend up to arbitrarily high masses. In this case, we can only place a lower bound on the WIMP mass and when calculating coverage statistics, we use 1-tailed limits (i.e. a \(p\%\) confidence limit encloses \(\frac{1}{2}(1+p) \%\) of the marginalised posterior).

We report coverage statistics for the various benchmark parameters in Table \ref{tab:CoverageAll}. For the Standard Halo Model, there is approximately exact coverage for both 50 and 100 GeV WIMPs, while for the VL-2 benchmark exact coverage is observed for the 100 GeV WIMP. The remaining benchmark parameters display some undercoverage, though still much improved over that achieved by the speed parametrisation method. The poorest coverage is achieved for the 100 GeV DD benchmark, for which the 68\% confidence interval has a coverage of \(58 \pm 3 \%\). This is to be expected from the poorly distributed reconstructions shown in Fig.\ \ref{fig:recons100}. For the 100 GeV dark disk benchmark, there appears to be a significant bias in the distribution of reconstructed values, which peaks around \(70 \textrm{ GeV}\). We explore the origin of this bias in the next section, where we examine the speed distributions reconstructed using this method.

\begin{table}[t]
  \begin{center}
\begin{ruledtabular}
    \begin{tabular}{lll}
    & \multicolumn{2}{c}{WIMP Mass} \\
     &  50 GeV & 100 GeV \\
  \hline
     \multirow{2}{*}{ \textsc{SHM}}  & \(71 \pm 3 \%\) & \(65\pm 3 \%\)\\
				&   \(92 \pm 2 \%\) & \(94 \pm 1 \%\)\\ \\
     \multirow{2}{*} {\textsc{DD}}  & \(61 \pm 3 \%\) & \(58 \pm 3 \%\) \\
				 & \(94 \pm 1 \%\) & \(91 \pm 2 \%\) \\ \\
     \multirow{2}{*} {\textsc{VL-2}} & \(72 \pm 3 \%\) & \(65 \pm 3 \%\)\\
				& \(90 \pm 2 \%\) & \(94 \pm 2 \%\)\\

    \end{tabular}
\end{ruledtabular}
  \end{center}
  \caption{68\% and 95\% confidence interval coverage results for the momentum parametrisation method using a variety of benchmark parameters, as defined in Sec.\ \ref{sec:ParamBenchmarks}.}
\label{tab:CoverageAll}
\end{table}

\subsection{Recovering the speed distribution}

We will now consider how the speed distribution can be reconstructed from the momentum parametrisation. For a set of constant bins in momentum space, the positions and widths of bins in velocity space is dependent on the WIMP mass. It is therefore difficult to extract precise statistical information on the speed distribution, as the bin values will be very strongly correlated with the WIMP mass. Instead, we take the reconstructed WIMP mass as fixed and use this to obtain a speed distribution from the momentum distribution parameters. Without treating the covariance of the WIMP mass and the bin parameters in full, the reconstructed speed distribution will depend strongly on the reconstructed mass value. However, this naive approach should give an indication of whether accurate reconstructions are possible.

\begin{figure}[t]
\centering
\includegraphics[width=0.5\textwidth]{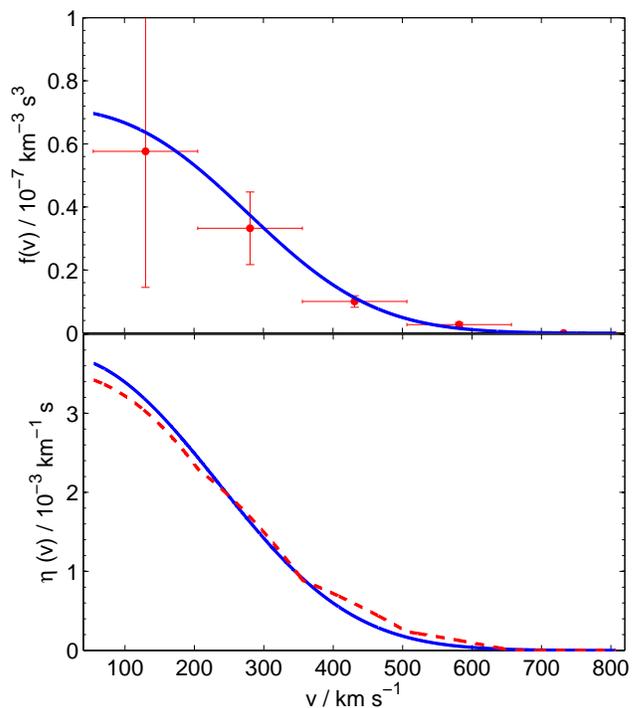}
\caption{Reconstructed speed distribution from all three mock experiments using the momentum parametrisation method. The benchmark is a 50 GeV WIMP and the SHM distribution function. The upper pane shows the underlying SHM speed distribution (solid blue) and the fitted values of the speed bin parameters (red points). The errors on the bin values are within-chain standard deviations as described in Sec.\ \ref{sec:ParamExplore}. The lower pane shows the mean inverse speed corresponding to these fitted values (dashed red line) and the true mean inverse speed (solid blue). The underlying distributions have been rescaled by \(\alpha\) for comparison to the reconstructions.}
  \label{fig:SHM50}
\end{figure}

First, we consider a 50 GeV WIMP with SHM distribution, as an archetypal WIMP model with a well-behaved distribution function. We show a typical reconstructed speed distribution in Fig.\ \ref{fig:SHM50}, using the same mock dataset as Fig.\ \ref{fig:50SHM_momentum}. In this case, the reconstructed value of \(m_\textrm{rec}\) is 42 GeV and the speed distribution appears to be accurately reconstructed within the error estimates. 

 \begin{figure}[t]
\centering
  \includegraphics[width=0.5\textwidth]{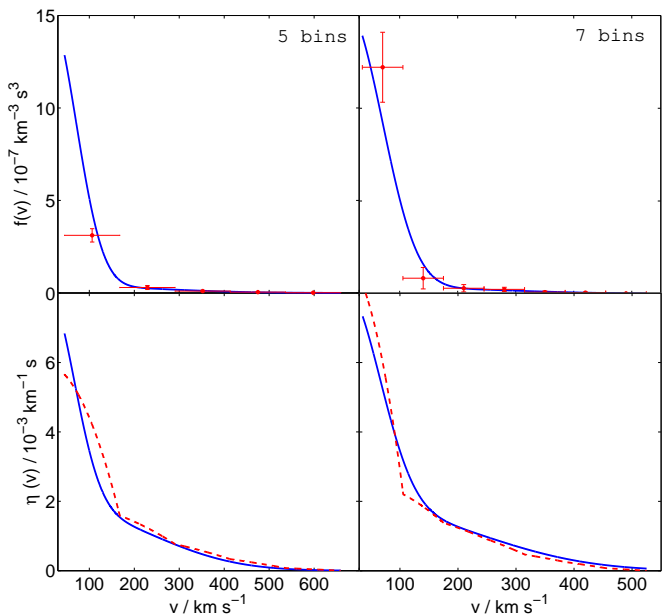}
\caption{As Fig.\ \ref{fig:SHM50} for a 100 GeV WIMP with DD distribution function using 5 momentum bins (left panes) and 7 momentum bins (right panes).}
  \label{fig:DD100}
\end{figure}

\begin{figure}[t]
\centering
\includegraphics[width=0.5\textwidth]{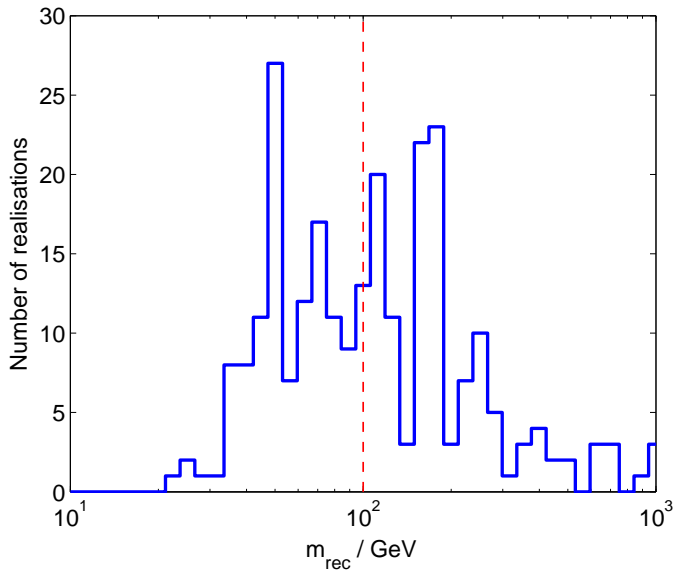}
\caption{Distribution of reconstructed masses using the 7-bin momentum method for 250 reconstructions for a DD benchmark distribution. The true mass of 100 GeV is shown as a dashed vertical line.}
  \label{fig:7bins}
\end{figure}

Next, we consider a reconstruction for a 100 GeV WIMP with DD distribution function. One example is shown in the left-hand panes of Fig.\ \ref{fig:DD100}, for a dataset with reconstructed mass \(\log_{10}(m_\textrm{rec} / \textrm{GeV}) = 1.83 \pm 0.15\), compared to the true value of \(\log_{10}(m_\chi / \textrm{GeV}) = 2\). The speed distribution appears to be well recovered at all speeds. However, there is a significant discrepancy in the mean inverse speed below \(\sim 150 \kms\). This is because the DD distribution function is very rapidly varying at low \(v\), meaning that the ansatz of constant bins can no longer be applied. As observed in the speed parametrisation method, the event spectrum can be steepened by moving to lower mass values and this may explain why there is significant bias and poor coverage for this set of benchmark parameters. 

In the right-hand panes of Fig.\ \ref{fig:DD100}, we show results from the same mock dataset reconstructed using 7 bins in momentum space. The reconstructed mass is now \(\log_{10}(m_\textrm{rec} / \textrm{GeV}) = 2.21 \pm 0.27\), with the mean inverse momentum more closely reconstructed than for the 5 bin case. Figure \ref{fig:7bins} shows the distribution of reconstructed masses for a 100 GeV WIMP with a DD distribution function using 7 bins in momentum space. The reconstructed masses are now more broadly distributed around the benchmark value, with improved coverage compared to the 5 bin case: \(67 \pm 3 \%\) and \(94 \pm 1 \%\). We have found that increasing the number of bins for the 50 GeV SHM benchmark leaves the coverage properties and distribution of reconstructions largely unchanged, indicating that increasing the number of bins can be used to check the robustness of the reconstructions.

\begin{figure}[t]
\centering
\includegraphics[width=0.5\textwidth]{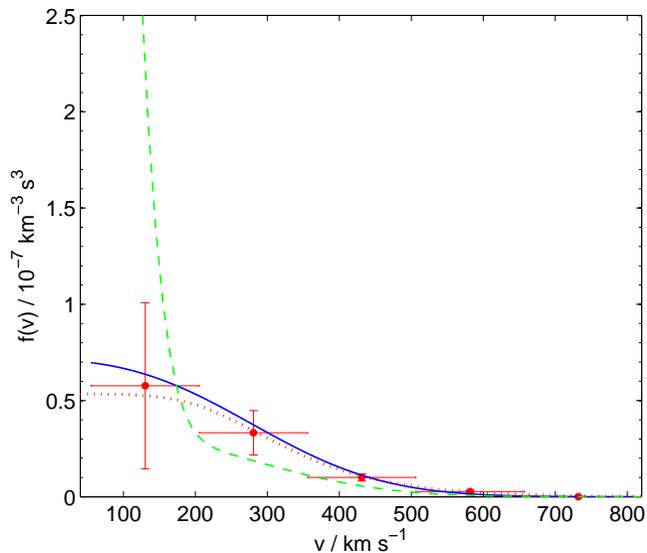}
\caption{Reconstructed speed distribution from all three mock experiments using a benchmark of a 50 GeV WIMP with SHM distribution. The reconstructed values have been rescaled by \(\alpha\) for comparison to the true distribution. The three different benchmark speed distributions defined in Sec.\ \ref{sec:ParamBenchmarks} have been overlaid: SHM (solid blue), DD (dashed green) and VL-2 (dotted red).}
  \label{fig:SHM50_all}
\end{figure}

Finally, we consider the discriminatory power of the reconstructions. Returning to the 50 GeV SHM benchmark, we plot a single speed distribution reconstruction in Fig.\ \ref{fig:SHM50_all}, as well as all three benchmark speed distributions for comparison. The reconstruction is reasonably consistent with both the SHM and VL-2 models and displays only mild tension with the DD model. In addition, the benchmark distributions in Fig.\ \ref{fig:SHM50_all} have been rescaled by the true value of \(\alpha\) for comparison with the reconstructed values. In a real experiment, the value of \(\alpha\) is unknown, further reducing the potential to discriminate between different models. Only in the case of more extreme distribution functions, such as a dark disk, might it be possible to make a distinction between the many possible underlying models. Thus, while the momentum parametrisation method can provide good constraints on the mass of the WIMP, it remains difficult to probe the speed distribution function.

\section{Conclusion}
\label{sec:Conclusion}
We have presented a novel method of analysing direct detection datasets which aims to reconstruct as much information as possible about the WIMP mass, cross-section and distribution function while making no assumptions about the shape of the underlying speed distribution. To do this, we parametrise the WIMP momentum distribution using a simple empirical parametrisation. This ensures that we do not parametrise those regions of velocity space to which the experiments are not sensitive, thus preventing spurious reconstructions, as seen previously in the use of speed parametrisation methods. 

In the case of a single experiment, this method can be applied exactly and allows one to extract information about the shape of the distribution function, at the cost of losing access to information about either the WIMP mass or cross-section separately. The overall `scale' parameter \(D\), defined in Eq.\ \ref{eq:D} which encodes information about both the mass and cross-section, can be reconstructed with only a small bias, and conservative limits can reliably be placed on \(D\). The \(D\) values from many different experiments can then potentially be used to place bounds on the values of the WIMP mass and cross-section. The momentum distribution is also accurately reconstructed, though an increased number of parameter bins may be required as the number of signal events increases.

 For multiple experiments, the range of the parametrisation must be extended to cover the sensitivity regions of all experiments. For estimation of the WIMP mass, this allows us to achieve significant improvements in coverage and reduction in bias over previous methods, though the momentum parametrisation method still suffers from some under-coverage for more extreme distribution functions, such as the dark disk. In these cases, the coverage properties can be improved by using more bins in momentum space, though at the cost of significantly broadening the distribution of reconstructed masses.   Without making any assumptions about the WIMP speed distribution, however, we cannot estimate the interaction cross-section due to its degeneracy with the fraction of WIMPs accessible to the experiments. This is an unavoidable problem, but is rendered somewhat irrelevant by large uncertainties in the local DM density.

Reconstruction of the WIMP speed distribution remains difficult. The finite sensitivity window of direct detection experiments means that information on the normalisation of \(f(v)\) is lost, making comparison to theoretical models difficult. At the event rates studied here, it does not appear to be possible to distinguish between different distribution functions. If a clear dark matter signal is observed at next-generation ton-scale detectors, however, it should still be possible to accurately estimate the WIMP mass in a model-independent fashion.

\begin{acknowledgments}
BJK and AMG are both supported by STFC. AMG also acknowledges support from the Leverhulme Trust. We are grateful to Juerg Diemand for access to the VL-2 data.
\end{acknowledgments}

\bibliographystyle{apsrev4-1}
\bibliography{bradkav}

\end{document}